\documentclass[prl,aps,showpacs,preprint]{revtex4}
\usepackage{color}
\usepackage{setspace}

\usepackage{amssymb}
\usepackage{graphicx}

\def\eqref#1{\mbox{(\ref{#1})}}

\begin{document}
\title{Visualization of Au Nanoparticles Buried in a Polymer Matrix\\
by Scanning Thermal Noise Microscopy}

\author{Atsushi Yao}
\affiliation{Department of Electronic Science and Engineering, Kyoto University, Katsura, Nishikyo, Kyoto, 615-8510, Japan}
\author{Kei Kobayashi} \email{keicoba@iic.kyoto-u.ac.jp} 
\affiliation{Department of Electronic Science and Engineering, Kyoto University, Katsura, Nishikyo, Kyoto, 615-8510, Japan}
\author{Kuniko Kimura} 
\affiliation{Department of Electronic Science and Engineering, Kyoto University, Katsura, Nishikyo, Kyoto, 615-8510, Japan}
\author{Shunta Nosaka} 
\affiliation{Department of Electronic Science and Engineering, Kyoto University, Katsura, Nishikyo, Kyoto, 615-8510, Japan}
\author{Hirofumi Yamada}
\affiliation{Department of Electronic Science and Engineering, Kyoto University, Katsura, Nishikyo, Kyoto, 615-8510, Japan}

\date{\today}

\begin{abstract}
We demonstrated visualization of Au nanoparticles buried 300 nm into a polymer matrix by measurement of the thermal noise spectrum of a microcantilever
with a tip in contact to the polymer surface.
The subsurface Au nanoparticles were detected as the variation in the contact stiffness and damping reflecting the viscoelastic properties of the polymer surface.
The variation in the contact stiffness well agreed with the effective
 stiffness of a simple one-dimensional model, 
which is consistent with the fact that the maximum depth range of the technique is far beyond the extent of the contact stress field.
\end{abstract}

\maketitle

Several researchers have recently demonstrated visualization of subsurface features with a nanometer-scale resolution
using various imaging schemes based on atomic force microscopy (AFM)~\cite{yamanaka1994ultrasonic,
bodiguel2004depth, shekhawat2005nanoscale,
tetard2008imaging, tetard2008elastic,
yamanaka2008evaluation,parlak2008contact,
shekhawat2009ultrasound, ashino2009revealing, 
tetard2010atomic, tetard2010new,tetard2011virtual,hu2011imaging,
killgore2011quantitative,kimura2013imaging,
ebeling2013visualizing, verbiest2013subsurface, vitry2014advances, verbiest2015beating}.
As the maximum depth range of the technique reaches on the order of one
micrometer and the potential applications include those in the industrial, biological and medical research fields, much attention has been paid to these techniques.
However, the imaging mechanisms and underlying physics are still not 
well understood. 
This is partly because all the schemes used for subsurface imaging
require excitation of the oscillation of the cantilever and/or sample
surface, and 
the key factors contributing to the subsurface contrasts could vary depending on the imaging schemes.

One of the major imaging schemes is to excite two piezoelectric
actuators located at the cantilever base and the bottom of the sample at
two different frequencies and detect the flexural oscillation of the
cantilever at the beat frequency, which is caused by the nonlinear
tip-sample interaction~\cite{shekhawat2005nanoscale,tetard2008imaging,tetard2008elastic,
shekhawat2009ultrasound, tetard2010atomic, tetard2010new, tetard2011virtual}.
The beat frequency is tuned at the contact resonance frequency ($f_{\rm c}$) of the cantilever to enhance the contribution of the nonlinear coupling to the imaging mechanism as well as the signal-to-noise ratio.
The technique is
referred to as heterodyne force microscopy (HFM) or
scanning near-field ultrasound holography
(SNFUH). It has been applied to subsurface imaging for various sample systems with a depth range of a few hundred nm,
but mainly for buried hard objects in a soft matrix.
The imaging mechanism by this scheme has been explained as the amplitude and phase modulation of the surface acoustic standing wave resulting from the interference of the ultrasound waves transmitted through the sample and cantilever~\cite{shekhawat2005nanoscale,verbiest2012subsurface}.

Another possible imaging scheme is to excite a piezoelectric actuator located
at the bottom of the sample at a frequency close to $f_{\rm c}$ and detect the flexural oscillation of the cantilever at $f_{\rm c}$.
The technique is referred to as 
atomic force acoustic microscopy (AFAM)~\cite{rabe1996vibrations,rabe2006atomic}.
Several researchers also reported visualization of subsurface features with a depth range of a few hundred nm by AFAM, of whose imaging mechanism was explained by the modulation of the contact stiffness due to the subsurface features~\cite{parlak2008contact,striegler2011detection}. They also found that the result was consistent with a finite element analysis.

We have also demonstrated visualization of Au nanoparticles buried 900 nm in a polymer matrix using HFM and AFAM~\cite{kimura2013imaging}.
We recently measured the contact resonance spectra of the cantilever while the tip was scanned over the surface by sweeping the frequency of the sample excitation at each pixel. Since the contact resonance spectra were not skewed, they were well fitted by the simple harmonic oscillator (SHO) model.
We found that the contact resonance spectrum was affected by the Au
nanoparticle underneath, and we concluded that the variation in the
contact stiffness and damping was playing a major role in making
subsurface contrasts in the AFAM images, while the tip-sample nonlinearity does not seem to significantly contribute~\cite{kimura2016}.

As already mentioned, most of the subsurface imaging experiments have
been based 
on the detection of the cantilever oscillation close to $f_{\rm c}$. 
We now believe that the contact resonance is playing a major role 
in producing the subsurface contrasts in the AFAM, at least for the solid
nanoparticles buried in a soft matrix. 
We then raised the question; do we really need to excite a cantilever
oscillation? 
If we just need to measure the contact resonance spectra on 
the surface for subsurface imaging, it should be possible to 
do the same thing with the thermal drive of the cantilever. This was the motivation of the study.

There have been several reports about the AFM measurements of the tip-sample interactions using thermally driven cantilevers~\cite{benmouna2004viscoelasticity,vairac2003towards, benmouna2003nanoscale, gelbert1999viscoelastic,roters1996distance, cleveland1995probing}.
Motivated by the above question and inspired by these previous studies,
we measured the thermal noise spectra of a cantilever on the polymer matrix with buried Au nanoparticles by introducing scanning thermal noise microscopy (STNM), which simply collects the contact resonance spectra of the cantilever at each pixel while the tip was scanned over the surface.

In this letter, we demonstrate visualization of the Au nanoparticles
buried 300 nm in a polymer matrix by STNM. 
Since STNM does not require additional excitation methods of the
cantilever base or sample surface, 
the contact resonance spectrum measured by STNM is free from the
spurious peaks that hinder quantitative estimation of the contact
stiffness and damping, and the nonlinear tip-sample interaction can be
minimized because of a very small oscillation amplitude. 
We quantitatively evaluated the differences in the contact stiffness and damping of the polymer surface areas with and without the Au nanoparticle underneath using a linear spring dashpot model,
and discuss the imaging mechanisms by STNM as well as those by other schemes.

\begin{figure}
\begin{center}
\includegraphics[width=1\linewidth]{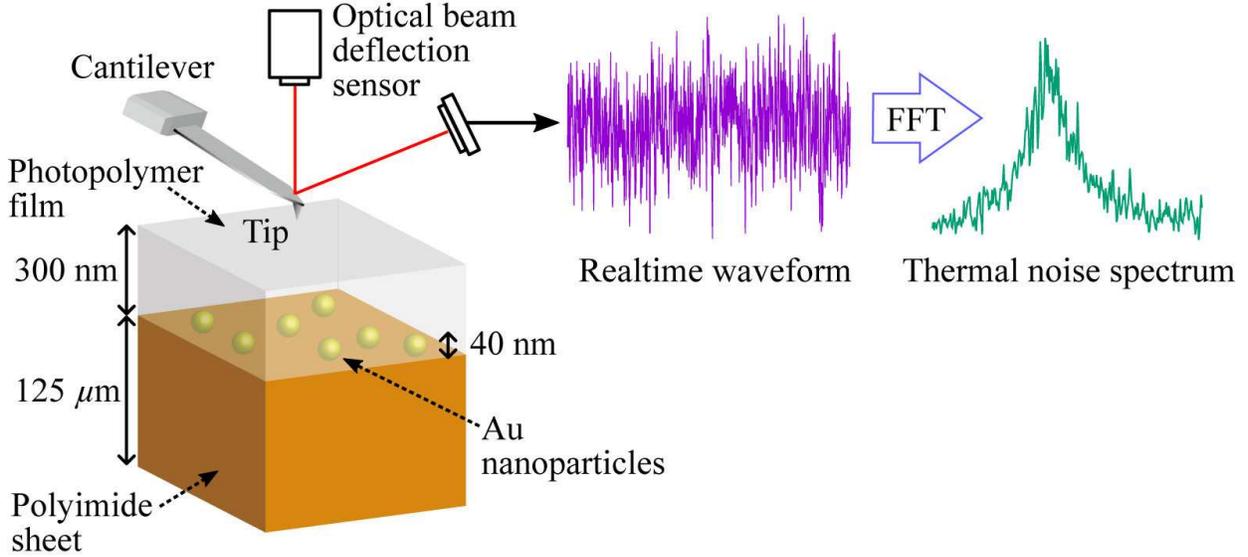}
\caption{Schematic of sample structure and experimental setup of scanning thermal noise microscopy (STNM).
Au nanoparticles were deposited on a polyimide sheet, which were subsequently covered with a 300-nm-thick photopolymer film (see Ref.~\cite{kimura2013imaging} for more details).
While the tip was scanning the surface with a constant loading force,
a realtime waveform of the cantilever deflection was recorded at each pixel.
The thermal noise spectrum was calculated by the fast Fourier transform algorithm (FFT).}
\label{fig1}
\end{center}
\end{figure}

Figure 1 shows a schematic of a sample's structure and the experimental
setup of the STNM.
We used a model sample of Au nanoparticles buried in a polymer matrix~\cite{kimura2013imaging}.
Au nanoparticles with diameters of 40 nm dispersed in water with a concentration of $0.006$--$0.007$ wt\% (Tanaka Kikinzoku Kogyo) were dropped onto a
125-$\mu$m-thick polyimide sheet (DuPont--Toray: Kapton 500V).
The sheet was dried on a hot plate heated at 105$^\circ$C. A photopolymer (Rohm and Haas: S1813G) was spin-coated as the top-coat and the sheet was annealed at
150$^\circ$C for 5 min. 
The thickness of the top-coat layer was determined by a stylus profiler (KLA--Tencor: P--15) to be about 300 nm.
More details about the sample preparation procedures and a cross-sectional scanning electron micrograph were published in Ref.~\cite{kimura2013imaging}.

We used a commercial AFM (JEOL: SPM 5200) after some modifications to the
optics and electronics to reduce the sensor noise in the optical beam deflection sensor~\cite{fukuma2005development}.
A multifunction data acquisition device (National Instruments: NI USB--6366) was used to acquire the realtime waveform from the deflection sensor, and the thermal noise spectrum was computed by the fast Fourier transform (FFT) algorithm.

We used a Si cantilever with a backside Al coating (Nanosensors: PPP-ZEILR).
We first measured the thermal noise spectrum of the first free resonance in air and it was fitted to the SHO model~\cite{Saulson,fukuma2005development},
to determine the first free resonance frequency ($f_0$ = 26.1 kHz) and the quality factor ($Q_0$ = 140), from which the spring constant of the cantilever ($k_z$) was calibrated by Sader's method to be 1.2 N/m~\cite{Sader}. We also calibrated the angular deflection sensitivity of the optical beam deflection sensor (See Supplementary Information A).

We brought the tip into contact with the sample surface at a loading force of 10 nN, and performed AFAM imaging at the first contact resonance
using a piezoelectric plate glued to the polyimide sheet and a lock-in amplifier (Zurich Instruments: HF2LI), and found some subsurface Au nanoparticle features (See Supplementary Information B for the details of the AFAM imaging).
We then performed STNM imaging on the same area. While the tip was scanning the surface, a realtime waveform of the cantilever deflection was recorded at each pixel for 625 ms with a sampling frequency of 400 kHz.
The waveform consisting of 250,000 data points was divided into 25 segments of 10,000 data points each. The thermal noise spectrum was calculated from each segmented waveform by the FFT algorithm, and the averaged thermal noise spectrum was obtained. The frequency resolution was 40 Hz. The total data acquisition time for two STNM images (trace and retrace) with 128 $\times$ 128 pixels each was about 6 h.

\begin{figure}[!t]
\begin{center}
\includegraphics[height=6.5cm]{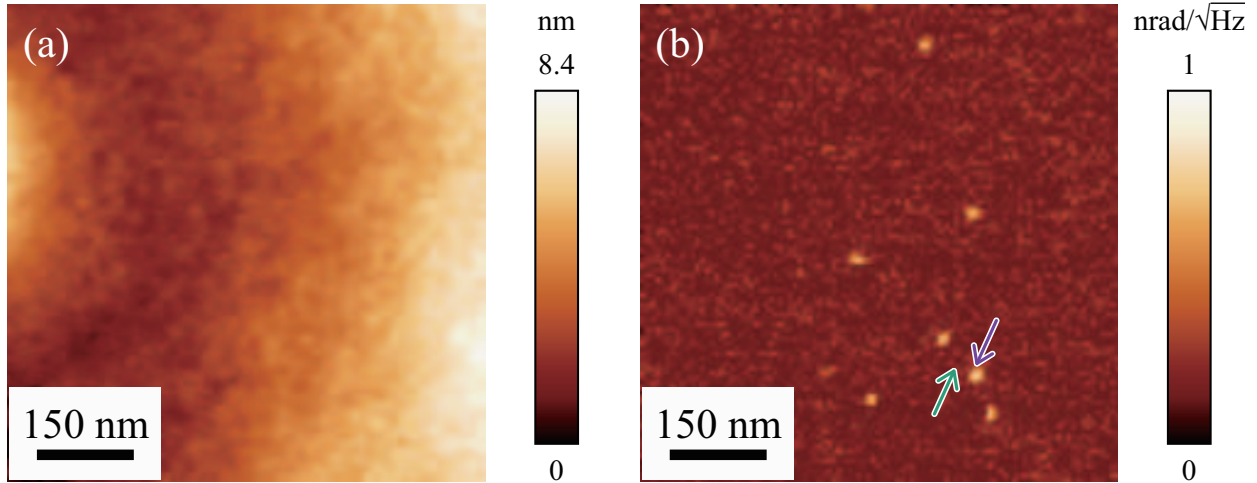}
\caption{(color online) STNM images of photopolymer film with Au
 nanoparticles buried at a depth of 300 nm.
The images are shown after trimming of an area of 780 nm $\times$ 780 nm (100 $\times$ 100 pixels.
(a) Topographic image showing a featureless photopolymer surface.
(b) Noise magnitude image reconstructed at 104.2 kHz showing buried Au nanoparticles as bright spots. The thermal noise spectra recorded at the locations indicated by the arrows are shown in Fig. 3.
}
\label{fig2}
\end{center}
\end{figure}

Figure 2(a) is a topographic image, which was obtained during STNM 
measurement, 
showing a smooth featureless surface of the top-coat layer.
Since we collected the thermal noise spectrum at each pixel, we can reconstruct the STNM noise magnitude image of an arbitrary frequency in the frequency range of concern.
Figure 2(b) is an STNM noise magnitude image at 104.2 kHz,
which shows the bright features at the same locations as those in the AFAM phase image (See Supplementary Information B).
The figure clearly shows well-dispersed Au nanoparticles buried 300 nm into the polymer matrix, as well as those presented in the previous papers\cite{shekhawat2005nanoscale, kimura2013imaging, kimura2016}, while there are no features at the same locations in the topographic image in Fig. 2(a).

\begin{figure}[!t]
\begin{center}
\includegraphics[height=7cm]{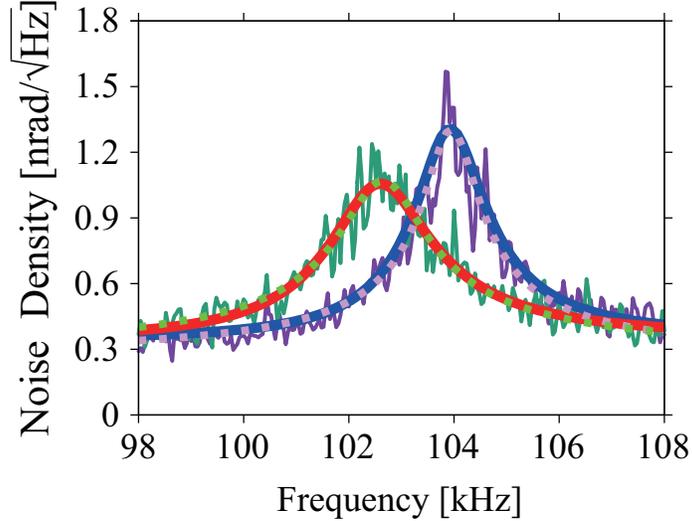}
\caption{(color online) Thermal noise spectra recorded on the
 photopolymer surface on an area with (purple) and without (green) a
 buried Au nanoparticle. The dashed and solid curves are fitted theoretical curves using Eqs. (1) and (3), respectively.
}
\label{fig3} 
\end{center}
\end{figure}

Figure 3 shows the thermal noise spectra measured on the pixels
indicated by the arrows in Fig. 2(b); the purple and green curves are
the thermal noise spectra with and without the buried Au nanoparticle
underneath, respectively. 
Unlike AFAM and other conventional techniques that utilize the piezoelectric actuator for excitation, the thermal noise spectrum measured by STNM is free from spurious resonance peaks and skewness caused by the nonlinear oscillations.
We again fitted the thermal noise spectra by
the SHO model,
\begin{equation}
N_{\rm{\theta}}^{f_{\rm c}} =
\sqrt {\frac{P_{\rm peak}}
{{Q_{\rm c}}^2[1-(f/f_{\rm c})^2]^2+(f/f_{\rm c})^2}+{n_\theta}^2},
\label{eq1}
\end{equation}
to determine the contact resonance frequency ($f_{\rm c}$) and the quality factor ($Q_{\rm c}$). $P_{\rm peak}$ is a fitting parameter corresponding to the peak noise power density of the angular deflection of the cantilever.
As shown in Fig. 3, the thermal noise spectra were well fitted by
Eq. (1) as the dashed curves, and we found that $f_{\rm c}$ and $Q_{\rm c}$ on the area with the Au nanoparticle were shifted to about 104.0 kHz and 77 from the values on the area without it, which were about 102.7 kHz and 53, respectively.
We calculated $f_{\rm c}$ and $Q_{\rm c}$ using the same method 
for all the thermal noise spectra, from which we reconstructed the $f_{\rm c}$ and $Q_{\rm c}$ images as shown in Figs. 4(a) and 4(b), respectively.
We can now see the bright features that represent the subsurface Au
nanoparticles both in the $f_{\rm c}$ and $Q_{\rm c}$ images, as clearly 
as in Fig. 2(b). Roughly speaking, $f_{\rm c}$ and $Q_{\rm c}$ are related to the contact stiffness and inverse of the damping.

Since STNM does not require external excitation methods, the subsurface contrasts are not contributed by the surface acoustic standing wave. The variation in the surface viscoelastic properties, namely, the contact stiffness and damping, should play a significant role.
In the following section, we assess the difference between the surface viscoelastic properties on the area with a buried Au nanoparticle and those on the area without it.
\begin{figure}[!t]
\begin{center}
\includegraphics[height=6.5cm]{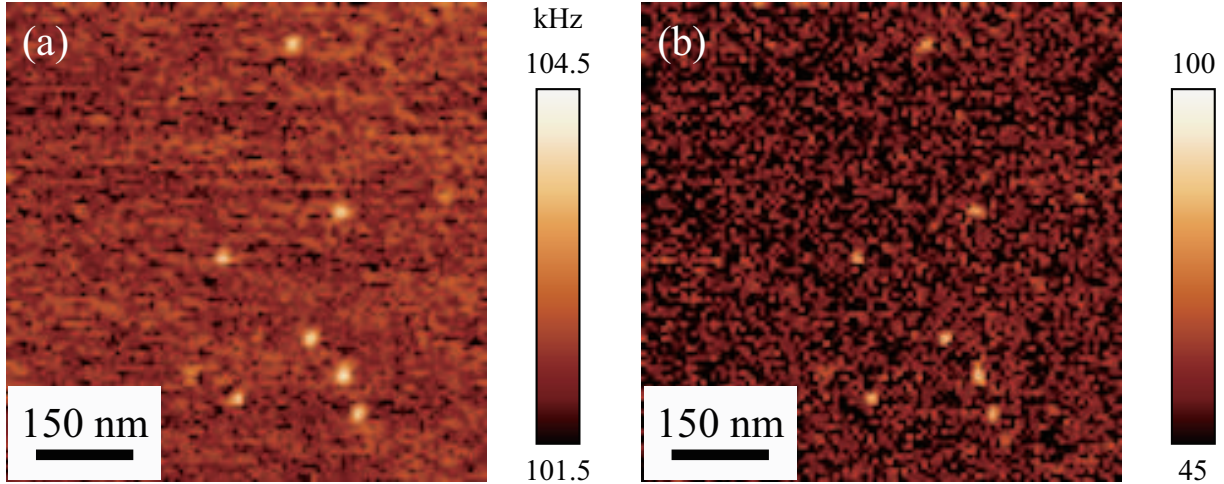}
\caption{(color online) (a) Contact resonance frequency ($f_{\rm c}$) and (b) quality factor ($Q_{\rm c}$) images of the photopolymer film with buried Au nanoparticles. Bright features correspond to the Au nanoparticles buried 300 nm below the photopolymer surface.
}
\label{fig4}
\end{center}
\end{figure}

We analyzed the thermal noise spectra using a linear spring dashpot model, i.e.,
the cantilever end is connected to the sample surface with a
spring of $k^*$ in parallel with a dashpot with damping $\gamma$
(Voigt model, See Supplementary Information
C)~\cite{rabe2006atomic,yuya2008contact}. 
We derived a fitting function for the thermal noise spectrum based on the frequency response function of the cantilever under
the boundary conditions for the AFAM, in which the sample surface is
excited, to determine $k^*$ and $\gamma$ (See Supplementary Information
C). 
We assume the thermal noise magnitude at the contact resonance is
proportional to the angular deflection at the cantilever end induced by
a unit sample surface oscillation, that is, a frequency response function for the angular deflection is given by
\begin{equation}
\theta_{z}^{\rm STNM}(\kappa)
=
\kappa \phi (\kappa)\frac{\sin (\kappa L) \sinh (\kappa L)}{N(\kappa)},
\label{eq2}
\end{equation}
where $\kappa$ and $L$ are a wave number and the length of the cantilever, respectively.
$\kappa$ is related to the oscillation frequency ($f$)
by $\kappa = \kappa_1 \sqrt[4]{ ({{f}/f_{0}})^2 - i f/ (f_{0} Q_{0})}$, where
$\kappa_1$, $f_{0}$, and $Q_{0}$ are the wave number (= 1.8751/$L$),
frequency, and quality factor of the first free resonance. 
$\phi(\kappa)$ and $N(\kappa)$ are given by
$\phi (\kappa) =3[k^* + i (2 \pi f_0)(\kappa/\kappa_1)^2 \gamma]/k_z$
and $N(\kappa)={(\kappa L)^3}( {1 + \cos \kappa L\cosh \kappa L}) - \phi (\kappa  )( {\cos \kappa L\sinh \kappa L-
\sin \kappa L\cosh \kappa L})$, respectively.
We derived the fitting function for the thermal noise spectrum obtained
by the STNM
as
\begin{equation}
N_\theta^{f_{\rm c}}= \sqrt{
\left|\theta_{z}^{\rm STNM}(\kappa) u_{\rm th}\right|^2
+{n_{\theta}}^2},
\label{eq3}
\end{equation}
where $u_{\rm th}$ is a fitting parameter corresponding to the thermal noise displacement at the cantilever end.
We fitted the thermal noise spectra in Fig. 3 with Eq. (3).
The red and blue curves in Fig. 3 show the best fitted curves 
to the measured spectra (green and purple curves). 
Based on the fitting parameters, $k^*$ and $\gamma$ on the area above the
Au nanoparticle were 63 N/m and 5.1 $\times 10^{-6}$ Ns/m, respectively,
while those on the area without it were 55 N/m and 6.1 $\times
10^{-6}$ Ns/m, respectively.
Therefore, $k^*$ was increased by 15 \% and $\gamma$ was decreased
by about 16 \% due to the presence of the Au nanoparticle in the matrix. 
Thus the imaging mechanism of the Au nanoparticles by STNM 
is quantitatively explained by the increase in the contact stiffness 
and damping of the polymer surface due to the existence of the Au
nanoparticle. 
Based on the fitting parameters, the magnitude of
the thermal displacement of the tip can also be estimated as
about $3.3$ pm and $4.2$ pm on the area with and without the buried Au
nanoparticle, respectively (See Supplementary information D). 
Note that we also calculated $k^*$ and $\gamma$ using another fitting
function based on 
the other boundary condition; i.e., 
a concentrated force is applied at the cantilever end, and obtained almost the same results.

We now discuss why the Au nanoparticles buried at such a depth can influence the surface stiffness and damping and eventually change the boundary conditions.
We considered two extreme cases; the sphere-plane contact models~\cite{johnson1987contact} and one-dimensional model to calculate the variation in $k^*$.
We first considered the AFM cantilever tip 
in contact with an elastic surface in which $k^*$ is defined as $dF_{\rm
n}/d\delta$, 
where $\delta$ and $F_{\rm n}$ are the surface displacement and normal loading force, respectively.
Under the Hertzian model (no adhesion), 
$\delta_{\rm Hertz}$ is given by $a_{\rm Hertz}^2/R_{\rm t}$, where $a_{\rm Hertz}$ is the contact tip radius given by $a_{\rm Hertz}=[3 R_{\rm t} F_{\rm n}/(4 E^*)]^{1/3}$
where $R_{\rm t}$ and $E^{*}$ are the tip radius and reduced Young's
modulus, respectively. 
Using these relationships, $E^*$ is related to $k^*$ as $E^*=\sqrt{k^{*3}/(6 R_{\rm t} {F_{\rm n}})}$.
By assuming $R_{\rm t}$ = 15 nm and $F_{\rm n}$ =10 nN,
$E^*$ on the area with and without the Au nanoparticle were calculated
as 16.7 GPa and 13.5 GPa, respectively, from the $k^*$ value obtained by
the STNM measurement. 
$E^*$ is related to the effective sample stiffness ($E_{\rm s})$ by $E^* = [(1 - \nu_{\rm t}^2)/E_{\rm t} + (1-\nu_{\rm s}^2)/E_{\rm s}]^{-1}$, 
where $E_{\rm t}$ denotes the Young's moduli of the tip (= 130
GPa)\cite{rabe1996vibrations}, and $\nu_{\rm t}$ (=
0.18)~\cite{rabe1996vibrations} and $\nu_{\rm s}$(=
0.33)~\cite{calabri2007nanoindentation} are the Poisson's ratios of the
tip and sample, respectively. Therefore, $E_{\rm s}$ value on the area
with and without the Au nanoparticle were calculated to be 16.7 GPa and 13.5 GPa, respectively.
These values are much larger than the Young's modulus of the top-coat photopolymer film in the literature~\cite{calabri2007nanoindentation} and that experimentally determined ($E_{\rm tc}$ = 3.4 GPa) (See Supplementary information E).
Moreover, the Hertzian model predicts that the stress field extends to a depth of about $3 a_{\rm Hertz}$\cite{johnson1987contact,rabe2002imaging}. Since
$a_{\rm Hertz}$ was about 2 nm in the present case,
it is not expected that the effective sample stiffness is affected by the Au nanoparticle buried 300 nm from the surface based on the Hertzian contact model.

We also considered the Johnson-Kendall-Roberts (JKR) contact model\cite{johnson1987contact}, in which the adhesion force was taken into account.
In the JKR model, $\delta_{\rm JKR}$ and $a_{\rm JKR}*$ are given by
$\delta_{\rm JKR}={a_{\rm JKR}*}^2/R_{\rm t}-2\sqrt{a_{\rm JKR}* F_{\rm ad}/(R_{\rm t} E^*)}$ and
$a_{\rm JKR}*=[3 R_{\rm t}\left(F_{\rm n}+2F_{\rm ad} + \sqrt{4F_{\rm n} F_{\rm ad} +4{F_{\rm ad}}^2}\right)/(4 E^*)]^{1/3}$, respectively,
where $F_{\rm ad}$ (= 15 nN) is the adhesion force.
Based on the JKR model,
$E^*$ on the area with and without 
the Au nanoparticle were calculated to be 11 GPa and 9 GPa,
respectively, from which the $E_{s}$ values were
10.6 GPa and 8.6 GPa, respectively.
This calculation suggests that the Young's modulus on the top-coat photopolymer was increased by about 23\%.
Note that these values are close to the Young's modulus of the top-coat photopolymer film in the literature~\cite{calabri2007nanoindentation}, but still greater than the experimental value.
Although we consider that the contact condition in the present study 
is more correctly described by the JKR contact model than by the Hertzian model, $a_{\rm JKR}$* by the JKR model was still as low as about 4.5 nm, and it does not account for the variation in the contact stiffness by the deeply buried Au nanoparticle.

Although the JKR model qualitatively explained the contact stiffness variation,
it failed to explain the depth range of the subsurface imaging
since the expected elastic stress field was on the order of 10 nm.
We finally noted that we found that the effective Young's modulus could be reproduced by considering a one-dimensional model.
We modeled the sample as a two-layer film of the top-coat layer and the Au layer. The variation in thickness of the two-layer film ($\delta_{\rm 1D}$) under a uniform stress ($\sigma$) is given by
$\delta_{\rm 1D}=(\sigma/E_{\rm tc})t_{\rm tc}+(\sigma/E_{\rm Au})t_{\rm Au}$, where
$t_{\rm tc}$, $t_{\rm Au}$, $E_{\rm tc}$, and $E_{\rm Au}$ 
denote the top-coat photopolymer thickness, Au nanoparticle diameter, and
the Young's modulus of the top-coat film and Au (= 79
GPa~\cite{dowson1990mechanics}), respectively. 
The effective Young's modulus of the multilayer film
can now be calculated as
$E_{\rm s}^{\rm 1D} = (\sigma/\delta_{\rm 1D})(t_{\rm tc} + t_{\rm Au})$.
Assuming $E_{\rm tc}$ as 3.4 GPa, the one-dimensional model predicts 
that Young's modulus on the top-coat photopolymer was increased by about
15\%, which was consistent with the variation predicted by the STNM measurements and the JKR contact model (23\%).
Therefore, we interpreted the subsurface contrasts in the experimental 
results by STNM as well as by AFAM as the variation in the contact
stiffness and damping because the stress field extends more than
expected from the sphere-plane contact models 
up to several hundreds of nm. This may be possible due to 
the anisotropy in the viscoelastic property of the spin-coated photopolymer film and some nonlinear or nanometer-scale effects that have not been considered in the macroscopic contact models.
The conclusion is also consistent with the previous studies that explained the subsurface imaging mechanisms using one-dimensional models~\cite{Sarioglu,cantrell2008analytical,cantrell2010}.
Further theoretical and experimental studies are necessary to comprehend the imaging mechanisms.

In conclusion, we experimentally performed the visualization of subsurface
features by STNM that simply measures the thermal noise spectrum without additional excitation methods.
We realized an ultimate simplification of the measurement scheme
and imaged the Au nanoparticles buried 300 nm in the photopolymer matrix in the least invasive way.
We have shown that the subsurface features in the STNM images were brought by the
variation in the contact stiffness and damping by the fitting of the
theoretical equation to the thermal noise spectra. 
It was also shown, based on the JKR contact model, that the Young's
modulus of the photopolymer surface was increased by 23\% due to 
the presence of the Au nanoparticle. However, the contact models failed
to explain 
the imaging mechanisms because the depth of the stress field in the model was more shallow than the depth of the Au nanoparticle.
On the other hand, the simple one-dimensional model also predicted
the increase in the Young's modulus of the photopolymer film by 15\%.
Therefore, we suggest that the imaging of subsurface features is
realized by the stress field extension in a quasi-one-dimensional manner.
To investigate the detailed one-dimensional strain model, 
the relationship between the effective Young's modulus of the top-coat polymer film and the top-coat thickness will be examined in the near future.
As shown by the STNM experiments, the subsurface features could be
solely 
explained by considering the variation in the viscoelastic properties of
the area under the tip. 
Therefore, the subsurface imaging mechanisms of various experimental schemes using the contact resonance might need to be revisited.
The STNM technique will serve as a technique to study subsurface imaging mechanisms by the AFM related techniques as well as a method to quantitatively evaluate the viscoelastic properties of the sample surface.

This study was supported by a Grant-in-Aid for JSPS Fellows
(No. 26462) and a Grant-in-Aid for Challenging Exploratory Research (No. 16K13686) from Japan Society for the Promotion of Science.

%

\pagebreak
\begin{center}
\textbf{\large {\rm Supplementary Information}\\
Visualization of Au Nanoparticles Buried in a Polymer Matrix\\ by Scanning Thermal Noise Microscopy}
\end{center}

\setcounter{equation}{0}
\setcounter{figure}{0}
\setcounter{table}{0}
\setcounter{page}{1}
\renewcommand{\theequation}{S\arabic{equation}}
\renewcommand{\thefigure}{S\arabic{figure}}
\renewcommand{\bibnumfmt}[1]{[S#1]}
\renewcommand{\citenumfont}[1]{S#1}

\section*{Supplementary Information A: Thermal Noise Spectrum of First Free Resonance}
Figure S1 shows a thermal noise spectrum of the first free resonance of the cantilever used for imaging of buried Au nanoparticles by STNM in the present study.
We fitted the measured thermal noise (voltage) spectrum ($N_{\rm v}^{f_0}$) to the equation for that of the simple harmonic oscillator (SHO)~\cite{SaulsonSI}, given by
\begin{equation}
N_{{\rm{v}}}^{{f_0}} = {S_z}N_{z}^{{f_0}} =
{S_z}\sqrt{
\frac{2k_{\rm B}T}{\pi f_0 k_z Q_0}
\frac{1}{[1-(f/f_0)^2]^2+[f/(f_0Q_0)]^2}+{n_z}^2},
\end{equation}
where $S_z$ and $n_z$ are a sensitivity of the optical beam deflection sensor to the displacement and a noise-equivalent displacement density, respectively. We first fitted the thermal noise spectrum to Eq. (S1) with the nominal value of $k_z$ and determined $f_0$ and $Q_0$, from which we calibrated $k_z$ by Sader's method~\cite{SaderSI}.
Then we fitted the spectrum again to Eq. (S1) with the calibrated $k_z$ to determine $S_z$ (= 25 mV/nm) and $n_z$ (= 77 fm/$\sqrt{\rm Hz}$)~\cite{HutterSI}.

Since the voltage output signal of the optical beam deflection sensor is proportional to the angular deflection frequency rather than the displacement of the cantilever, it is reasonable to rewrite Eq. (S1) as
\begin{equation}
N_{{\rm{v}}}^{{f_0}} = {S_{\theta}}N_{\theta}^{{f_0}} =
{S_{\theta}}\sqrt{
{\theta_z^{f_0}}^2
\frac{2k_{\rm B}T}{\pi f_0 k_z Q_0}
\frac{1}{[1-(f/f_0)^2]^2+[f/(f_0Q_0)]^2}+{n_\theta}^2},
\end{equation}
where $S_\theta$ and $n_\theta$ are a sensitivity of the optical beam deflection sensor to the angular deflection and a noise-equivalent angular deflection density, respectively.
$\theta_z^{f_0}$ is a conversion
factor of the displacement to the angular deflection at the first free resonance (= 1.3765/$L$), where $L$ denotes the length of the cantilever (= 450 $\mu$m).
Since $S_\theta$ is related to $S_z$ by $S_\theta = S_z/\theta_z^{f_0}$,
$S_\theta$ (= 8,170 V/rad) and $n_\theta$ (= 0.24 nrad/$\sqrt{\rm Hz}$)
were readily determined.

\begin{figure}[!h]
\begin{center}
\includegraphics[height=6.7cm]{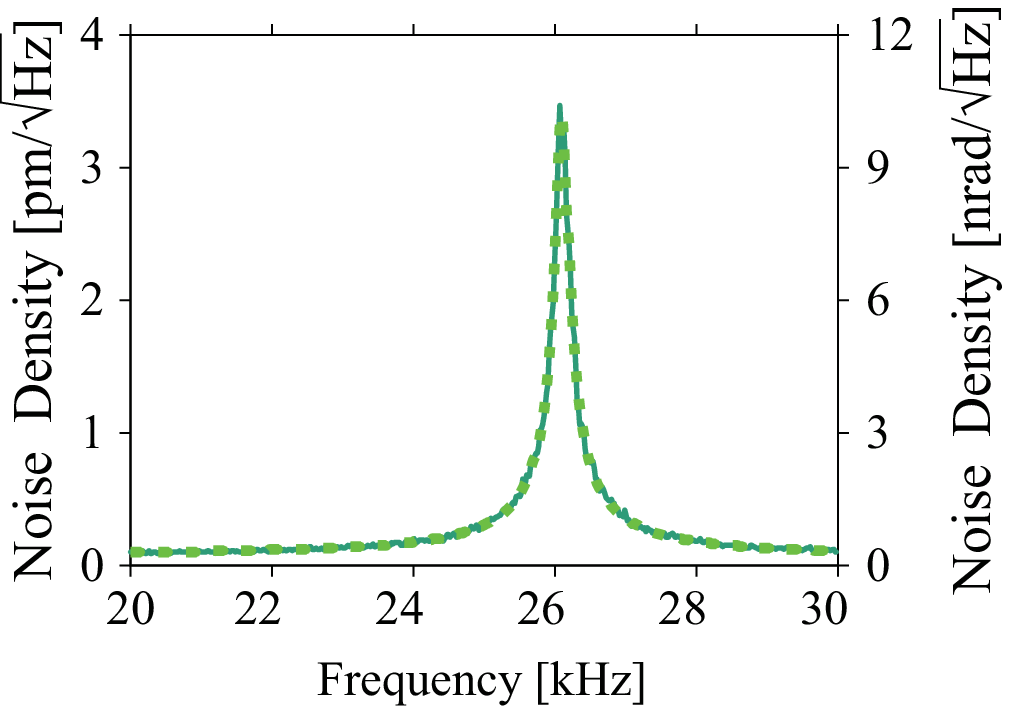}
\caption{Thermal noise spectrum of the first free resonance of the cantilever. The solid and dashed curves are the measured spectrum and the theoretical curve fitted with Eq. (S1), respectively.} 
\label{figS1}
\end{center}
\end{figure}

\section*{Supplementary Information B: AFAM Phase Imaging}
Figure S2 shows an atomic force acoustic microscopy (AFAM) phase image 
of a photopolymer film with Au nanoparticles buried 300 nm in depth. 
An excitation signal of 99.6 kHz with an amplitude of 8.5 mV was applied 
to a piezoelectric actuator glued to the backside of the polyimide
sheet. See Ref.~\cite{kimura2013imagingSI} for the details of the AFAM
experimental setup. 
The image area is the same as those of Figs. 2 and 4.

\begin{figure}[!h]
\begin{center}
\includegraphics[height=6.7cm]{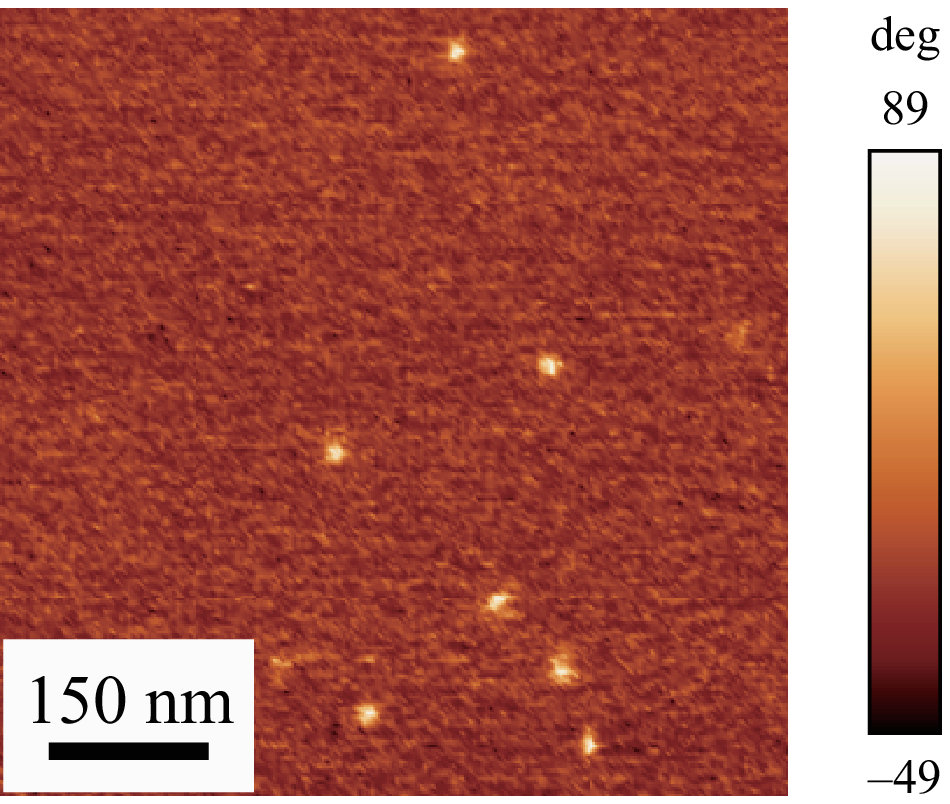}
\caption{AFAM phase image of a photopolymer film with Au nanoparticles buried 300 nm in depth.}
\label{figS2}
\end{center}
\end{figure}

\section*{Supplementary Information C: Fitting Function for Thermal
 Noise Spectrum of STNM}

We derive a fitting function for the thermal noise spectrum of the
contact resonance based on the frequency response function of AFAM. 
The tip-sample interaction is modeled by a linear spring dashpot model, i.e., the cantilever end is connected to the sample surface with a spring and a dashpot, representing the contact stiffness ($k^*$) and damping ($\gamma$), respectively, as shown in Fig. S3. The equation of motion for damped flexural oscillation of the cantilever is given by
\begin{equation}
EI\frac{\partial^4 y}{\partial x^4}+\eta\rho S
\frac{\partial y}{\partial t}+\rho S\frac{\partial^2y}{\partial x^2}=0
\end{equation}
where $E$ is the Young's modulus of the cantilever, $\rho$ is its mass density, $S$ is the area of its cross section, $I$ is the area moment of inertia, and $\eta$ is a damping constant.
$y$ represents the deflection of the cantilever in its thickness direction at the position of $x$, which is a
coordinate in length direction of the cantilever~\cite{rabe2006atomicSI}.
The mode shape function, $y(x)$, can be expressed as
\begin{equation}
y(x) = A_{1} (\cos \kappa x + \cosh \kappa x ) + A_{2} (\cos \kappa x -
  \cosh \kappa x ) + A_{3} (\sin  \kappa x + \sinh \kappa x )+ A_{4} (\sin  \kappa x - \sinh \kappa x )
\end{equation}
where $A_{1}$, $A_{2}$, $A_{3}$, and $A_{4}$ denote constants and $\kappa$ is a wave number. The boundary conditions for AFAM are 
\begin{eqnarray}
y(0) & = & 0 \ {\rm and}\\
y'(0) & = & 0
\end{eqnarray}
for the cantilever base ($x=0$), and
\begin{eqnarray}
y''(L) & = & 0 \ {\rm and} \\
y'''(L) & = & \frac{\phi (\kappa )}{L^{3}}\left[y(L) -u_{0}\right]
\end{eqnarray}
for the cantilever end ($x=L$), where $u_{0}$ is an oscillation amplitude of the sample surface.
$\phi (\kappa)$ is a contact function given by
\begin{equation}
\phi (\kappa) =\frac{3}{k_z}(k^*+i\omega \gamma)= 3\left[\frac{k^{*}}{k_z} + i(\kappa L )^{2} \frac{2 \pi \gamma f_{0} }{ (1.8751)^2 k_z }\right],
\end{equation}
where $\omega$ is an angular frequency (= 2$\pi f$).
By calculating the constants $A_1$, $A_2$, $A_3$, and $A_4$,
$y(x)$ is obtained as
\begin{eqnarray}
y(x) & = & - \frac{u_{0}}{2} \phi (\kappa ) \frac{\sin \kappa L + \sinh
 \kappa L}{N (\kappa)} (\cos \kappa x - \cosh \kappa x ) \nonumber \\ 
& +& \frac{u_{0}}{2} \phi (\kappa ) \frac{\cos \kappa L + \cosh
 \kappa L}{N (\kappa)} (\sin \kappa x - \sinh \kappa x ), 
\end{eqnarray}
with $N(\kappa)$ given by
\begin{equation}
N (\kappa) = (\kappa L)^{3} (1+ \cosh  \kappa L \cos 
  \kappa L ) -\phi(\kappa) (\sinh \kappa L \cos
  \kappa L - \cosh \kappa L \sin \kappa L).
\end{equation}
Since we use the optical beam deflection sensor, whose output signal is proportional to the angular deflection of the cantilever, we also calculate the derivative of the mode shape function as
\begin{eqnarray}
y'(x) &= & \kappa \frac{{{u_0}}}{2}\phi \left( \kappa
										\right)\frac{{\sin \kappa L +
\sinh \kappa L}}{{N\left( \kappa  \right)}}\left( { \sin \kappa x +
					\sinh \kappa x} \right) \nonumber \\
&+& \kappa \frac{{{u_0}}}{2}\phi \left( \kappa  \right)\frac{{\cos \kappa L + \cosh \kappa L}}{{N\left( \kappa  \right)}}\left( {\cos \kappa x - \cosh \kappa x} \right).
\end{eqnarray}
By substituting $x$ with $L$ for these equations, the oscillation amplitude of the displacement and the angular deflection at the cantilever end for AFAM are obtained as
\begin{equation}
y_{\rm AFAM}(L)= u_{0} \phi (\kappa) \frac{\sin \kappa L \cosh \kappa L
 - \sinh \kappa L \cos \kappa L}{N(\kappa)}  \label{Yafam}
\end{equation}
and
\begin{equation}
y_{\rm AFAM}'(L)= \kappa u_{0} \phi (\kappa) \frac{\sin \kappa L \sinh \kappa L }{N(\kappa)},
\end{equation}
respectively.
Here we define a frequency response function of the angular deflection for STNM, namely the angular deflection amplitude of the cantilever end induced by a unit sample surface oscillation as
\begin{equation}
\theta_{z}^{\rm STNM}(\kappa)=\frac{
y_{\rm AFAM}'(\kappa)}{u_0}=
\kappa \phi (\kappa)\frac{\sin (\kappa L) \sinh (\kappa L)}{N(\kappa)}. 
\end{equation}

\begin{figure}[!t]
\begin{center}
 \includegraphics[height=5.7cm]{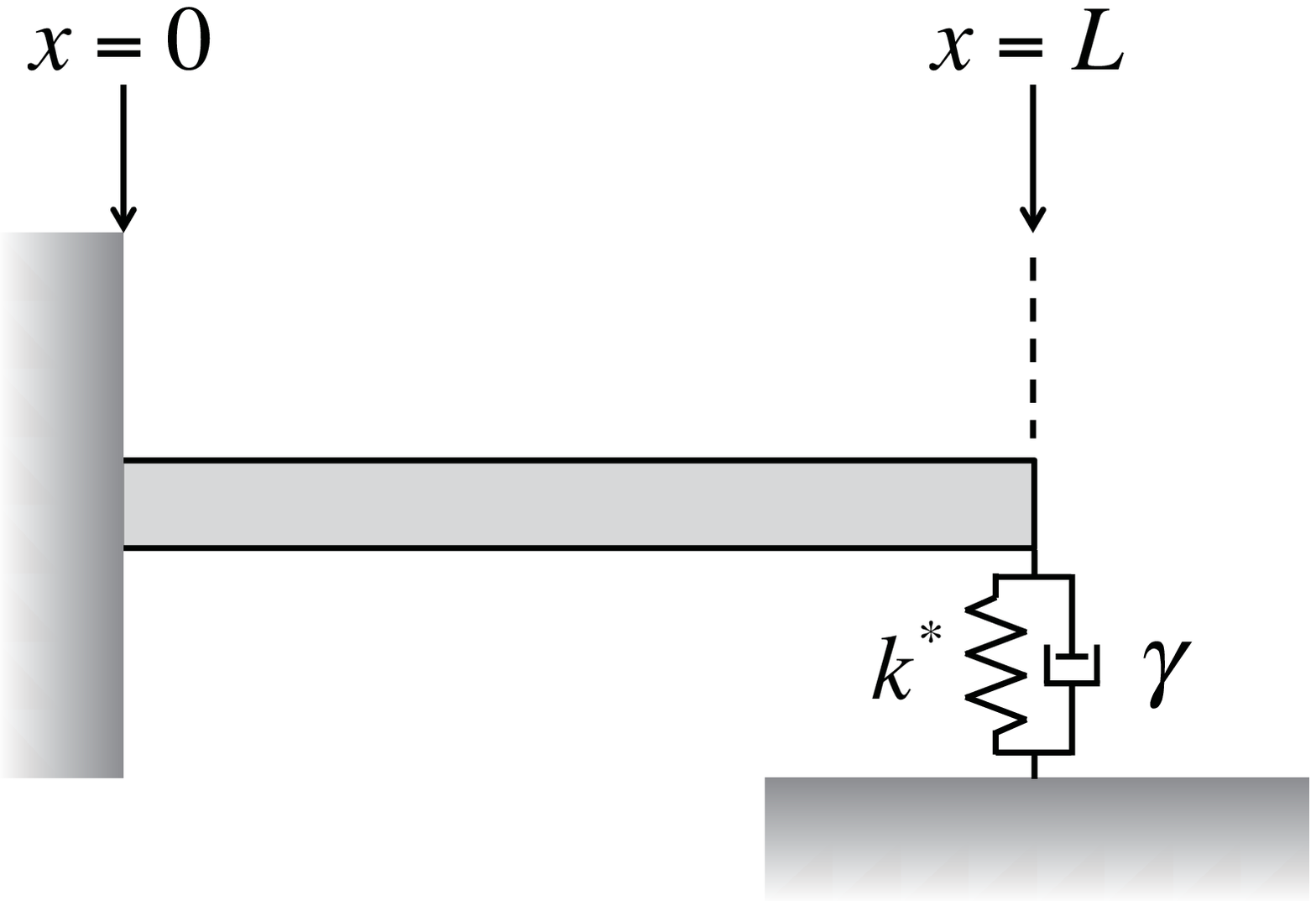}
\caption{Schematic of a cantilever interacting with a sample surface.
The tip-sample interaction is modeled by a linear spring dashpot model (Voigt model).}
\label{Bound}
\end{center}
\end{figure}

\section*{Supplementary Information D: Magnitude of Thermal Displacement of Tip}
\begin{figure}[!b]
\begin{center}
\includegraphics[height=5.2cm]{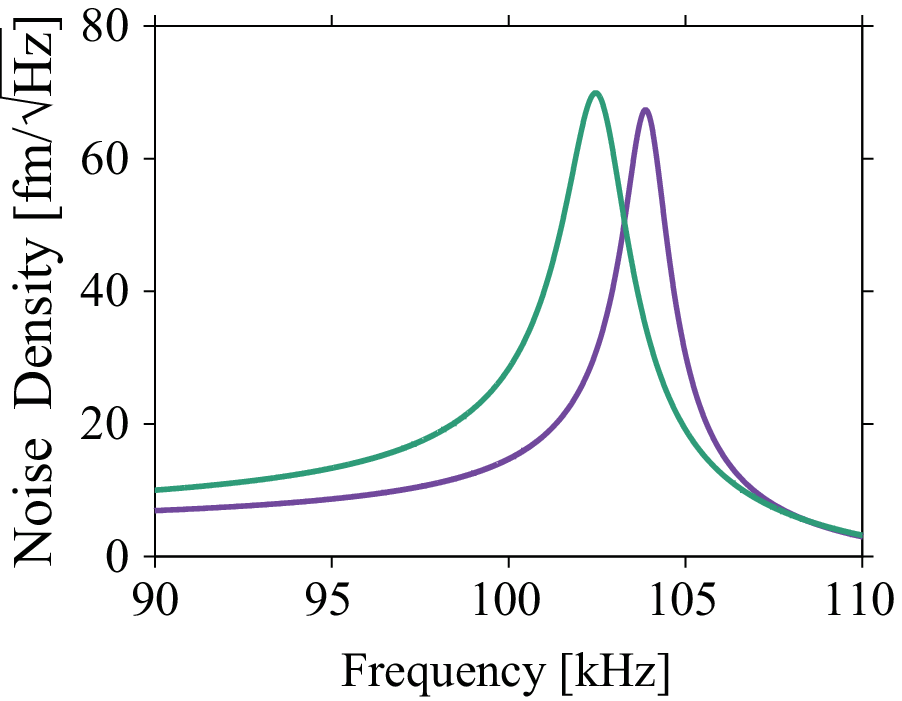}
\caption{(color online) Theoretical thermal noise spectrum of the
 tip calculated using the best fitting parameters for fitting Eq. (3) to
 the curves in Fig. 3. The purple and green curves show theoretical curves on
 the photopolymer surface areas with and without a buried Au
 nanoparticle using Eq. (S13).}
\label{FigS4}
\end{center}
\end{figure}

By inserting the best fitting parameters for $k^{*}$ and $\gamma$ to
Eq. (S13) and replacing $u_{0}$ with $u_{\rm th}$ obtained by the
fitting, we can recover the magnitude of the displacement density of 
the cantilever end (tip) as shown in Fig S4. 
The magnitude of the thermally driven tip oscillation at 
the first contact resonance was estimated as about $3.3$ pm and $4.2$ pm on 
the photopolymer surface areas 
with and without the buried Au
nanoparticle, respectively.

\section*{Supplementary Information E: Young's Modulus of Photopolymer Film}
We fabricated a patterned photopolymer film on the polyimide sheet, with a thickness of 250 nm, by photolithography, and performed STNM measurement at the edge of the film.
Fig. S5 shows the thermal noise spectra of the contact resonance of the
cantilever recorded on the areas on the photopolymer film and the
polyimide sheet. The purple and green curves are a typical thermal
noise spectrum on the photopolymer film and that on the polyimide sheet,
respectively. 
As we could not find a significant difference in the resonance frequency
or quality factor, 
we consider that the Young's modulus of the photopolymer film 
used in this study is almost the same as that of the polyimide sheet (= 3.4 GPa),
which is found in the literature~\cite{agesageSI}.

\begin{figure}[!h]
\begin{center}
\includegraphics[height=6cm]{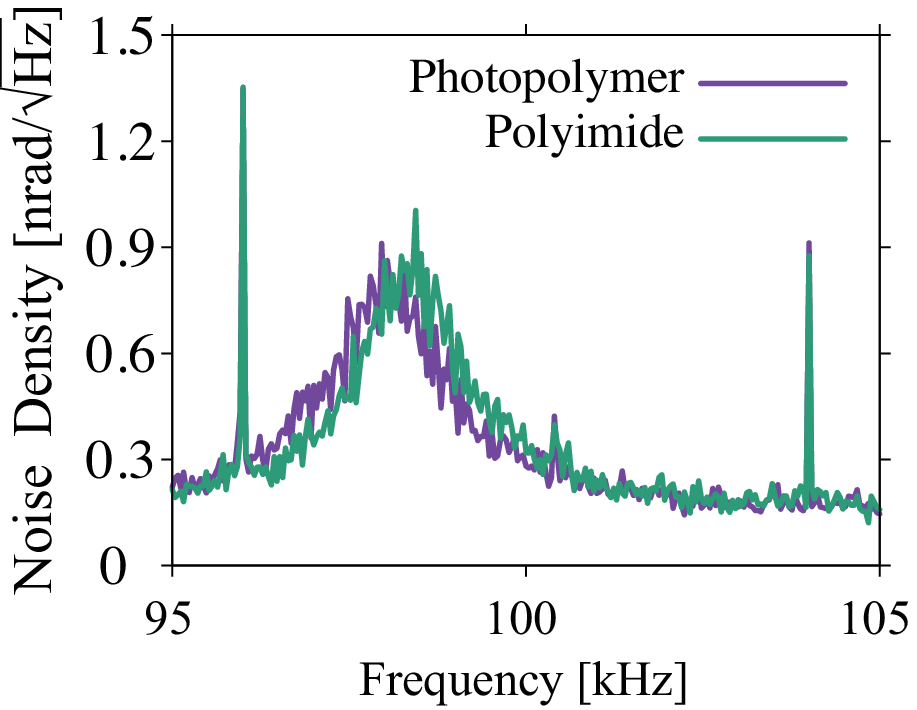}
\caption{(color online) Typical thermal
noise spectra recorded on the photopolymer film surface and the polyimide
sheet.}
\label{FigS5}
\end{center}
\end{figure}


%

\end{document}